\documentclass[aps,superscriptaddress,floatfix,showpacs,showkeys,preprintnumbers,amsmath,amssymb]{revtex4}
\usepackage[T1]{fontenc}
\usepackage[latin1]{inputenc}
\usepackage{graphicx}% Include figure files
\usepackage{dcolumn}% Align table columns on decimal point
\usepackage{color}
\usepackage[normalem]{ulem}

\newcommand{\be}{\begin{eqnarray}}
\newcommand{\ee}{\end{eqnarray}}
\newcommand{\ds}{\displaystyle}
\newcommand{\romq}{\mathrm q}

\begin{document}

\title{Continued fraction approximation for the nuclear matter response function}

\author{J. Margueron}
\affiliation{Institut de Physique Nucl\'eaire (UMR 8608), 
CNRS/IN2P3-Universit\'e Paris-Sud, F-91406 ORSAY CEDEX, France}
\author{J. Navarro}
\affiliation{IFIC (CSIC - Universidad de Valencia), Apdo. 22085,
E-46.071-Valencia, Spain}
\author{Nguyen Van Giai}
\affiliation{Institut de Physique Nucl\'eaire (UMR 8608), 
CNRS/IN2P3-Universit\'e Paris-Sud, F-91406 ORSAY CEDEX, France}
\author{P. Schuck}
\affiliation{Institut de Physique Nucl\'eaire (UMR 8608), 
CNRS/IN2P3-Universit\'e Paris-Sud, F-91406 ORSAY CEDEX, France}

\date{ \today}

\begin{abstract}
We use a continued fraction approximation to calculate the RPA
response function of nuclear matter. The convergence of the
approximation is assessed by comparing with the numerically exact
response function obtained with a typical effective finite-range
interaction used in nuclear physics. 
It is shown that just the first order term of the
expansion can give reliable results at densities up to the
saturation density value.
\end{abstract}



\maketitle

\section{Introduction}
There are many physical issues that require the knowledge of the
response function of a medium to an external probe. Well-known
examples are the electron scattering by nuclei or the propagation
of neutrinos in nuclear matter. In order to develop a microscopic
theory of response functions in finite nuclear systems one usually
starts by considering the limiting case of an infinite medium.
Infinite nuclear matter as a homogeneous medium made of
interacting nucleons is a very useful and broadly used concept
because of its relative simplicity and its connection with the
bulk part of atomic nuclei. A popular
approach 
consists in using an effective nucleon-nucleon interaction
adjusted to describe the nuclear matter properties
in a mean field approximation. Then, this microscopic description
can be extended to finite nuclei.

 In a mean field framework the nuclear
response functions must take into account the effects of
long-range correlations by the Random Phase Approximation (RPA)
which is the small amplitude limit of a time-dependent mean field
approach. This is well suited for those excitations which
correspond to small amplitude vibrations, the most typical of
which being the giant resonances and the low-lying collective
states~\cite{Harakeh}. For the theory to be consistent, it is
necessary that the same effective nucleon-nucleon interaction
generates the self-consistent Hartree-Fock (HF) mean field and the
RPA correlations which lead to the excitations of the system.

There are two types of interactions widely used in
non-relativistic approaches, the zero-range Skyrme-type
forces~\cite{Vau72} and the finite-range Gogny-type
forces~\cite{Gog75}. Skyrme forces are very often used because of
their relatively simple analytic form which allows for quite
complete RPA calculations in nuclear matter~\cite{Gar92} as well
as in finite nuclei~\cite{Terasaki,Matsuo}. On the other hand,
finite-range forces require heavier computational efforts to
calculate RPA responses 
in nuclei~\cite{Blaizot77,Peru05}. Furthermore, the only existing
methods in this case 
consist in diagonalizing large size matrices in configuration
space. It would be useful to have alternative methods such as a
direct calculation in coordinate space or momentum space of RPA
response functions, to avoid the increasingly large configuration
spaces. This is possible with Skyrme forces~\cite{Bertsch75} but
in the case of finite range forces the exchange interactions
complicate the problem.

In this work we study an approximation based on a continued
fraction expansion of the response function. Our aim is to explore
a calculational scheme which can be checked in infinite matter and
which offers prospects for RPA calculations with finite range
forces in nuclei. The continued fraction method is known in the
literature~\cite{sch89} and it has been used by many authors to
study response functions in the quasi-elastic regime (see
Ref.~\cite{Pace98} and references therein). However, it is
difficult to know where to truncate the continued fraction
expansion to obtain a desired accuracy. It is possible to
calculate response functions in infinite matter by performing
multipole expansions of the interaction and to have numerically
accurate results~\cite{mar05} to evaluate various approximation
schemes. Therefore, the present study aims at assessing the speed
of convergence of the continued fraction expansion applied to the
response functions in nuclear matter, using as an example a Gogny force
D1~\cite{Gog75}. We show that this expansion gives good results as
compared with the numerically exact calculations, even at lowest
order.

In Sec.II we recall the basic features of the continued fraction
method applied to the determination of RPA response functions in
an infinite medium, and we show analytically that it gives the
correct result in the special case of a Landau-Migdal interaction.
In Sec.III we discuss the results obtained with a finite range
interaction of Gogny type. Conclusions are drawn in Sec.IV~.

\section{Formalism}

\subsection{General framework}

A general two-body interaction in momentum representation depends
at most on 4 momenta. Because of momentum conservation there are
actually 3 independent momenta, in the case of a translationally
invariant interaction. For the particle-hole (p-h) case
we choose these independent variables to be the initial (final)
momentum ${\bf k_1}$ (${\bf k_2}$) of the hole and the external
momentum transfer ${\bf q}$. We follow the notations of
Ref.~\cite{mar05} and we denote by $\alpha = (S,M;T,Q)$ the spin
and isospin p-h channels with $S$=0 (1) for the non spin-flip
(spin-flip) channel, $T$=0 (1) the isoscalar (isovector) channel,
$M$ and $Q$ being the third components of $S$ and $T$. The matrix
element of the general p-h interaction including exchange can be
written as: 
\be 
&& V_{\rm ph}^{(\alpha,\alpha')}({\bf q},{\bf
k}_1,{\bf k}_2) \equiv \nonumber\\
&& \quad \langle {\bf q}+{\bf k}_1, {\bf k}_1^{-1},(\alpha) 
| V | {\bf q}+{\bf k}_2, {\bf k}_2^{-1},
(\alpha') \rangle~.
\label{Vph}
\ee

To calculate the response of a homogeneous medium to an external
field it is convenient to introduce the Green's function, or
retarded p-h propagator $G^{(\alpha)}({\bf q},\omega,{\bf k}_1)$.
From now on we choose the $z$ axis along the direction of ${\bf
q}$. In the HF approximation the p-h Green's function is the free
retarded p-h propagator~\cite{fet71}: 
\be 
G_{\rm HF}(\romq,\omega,{\bf k}_1) = \frac{f(k_1)-
f(\vert{\bf k}_1+{\bf q}\vert)} 
{\omega +
\epsilon(k_1) - \epsilon(\vert{\bf k}_1 +{\bf q}\vert) 
+ \mathrm{i} \eta}~,
\label{GHF} 
\ee 
where $\epsilon(k)$ is the HF single-particle
energy corresponding to momentum $\bf{k}$, and the Fermi-Dirac
distribution $f$ is defined for a given temperature $T$ and
chemical potential $\mu$ as
$f(k)=[1+e^{(\epsilon(k)-\mu)/T}]^{-1}$. The HF Green's function
$G_{\rm HF}$ does not depend on the spin-isospin channel $\alpha$.

To go beyond the HF mean field approximation one takes into
account the long-range type of correlations by  re-summing a class
of p-h diagrams. One thus obtains the well-known RPA~\cite{fet71}
whose correlated Green's function $G_{\rm
RPA}^{(\alpha)}(\romq,\omega,{\bf k}_1)$ satisfies the
Bethe-Salpeter equation: 
\begin{widetext}
\be 
G_{\rm
RPA}^{(\alpha)}(\romq,\omega,{\bf k}_1) = G_{\rm
HF}(\romq,\omega,{\bf k}_1) + G_{\rm HF}(\romq,\omega,{\bf k}_1)
\, \sum_{(\alpha')} \, \int \frac{{\rm d}^3k_2}{(2 \pi)^3} V_{\rm
ph}^{(\alpha,\alpha')}(\romq,{\bf k}_1,{\bf k}_2) G_{\rm
RPA}^{(\alpha')}(\romq,\omega,{\bf k}_2)~. \label{eqBS} 
\ee
\end{widetext}
Finally, the response function $\chi^{(\alpha)}(\romq,\omega)$ in
the infinite medium is related to the p-h Green's function by: 
\be
\chi_{\rm RPA}^{(\alpha)}(\romq,\omega) = g \int \frac{{\rm d}^3
k_1}{(2 \pi)^3} G_{\rm RPA}^{(\alpha)}(\romq,\omega,{\bf k}_1)~,
\label{chi} 
\ee 
where the spin-isospin degeneracy factor $g$ is 4
in symmetric nuclear matter and 2 in pure neutron matter. In the
case of a system of particles without residual interactions the
free response is obtained by calculating Eq.~(\ref{chi}) with the
 HF p-h propagator $G_{\rm HF}$, thus obtaining the
well-known Lindhard function $\chi_{\rm HF}$.

\subsection{Continued fraction approximation}

A direct numerical solution of Eq.~(\ref{eqBS}) with a general p-h
interaction is possible, as it has been shown in Ref.~\cite{mar05}
for the Gogny interaction. However, such a method is specifically
designed for infinite systems and it would be interesting to have
an alternative method which can be accurate and at the same time
can be used in calculations of finite systems. We examine now an
approximate way to calculate the RPA response function, expressing
it as a continued fraction. To simplify the writing of the
equations we shall employ the following conventions. First of all
we omit the variables such as $\romq$, $\omega$ or ${\bf k}$ as well
as indices $(\alpha)$, unless necessary. For instance,
equation~(\ref{eqBS}) is written as 
\be 
G_{\rm RPA} = G_{\rm HF} +
G_{\rm HF} V_{\rm ph} G_{\rm RPA}~. \label{eqBS2} 
\ee 
Secondly,
for any function $F({\bf k}_1)$ depending on a momentum ${\bf
k}_1$ we denote by $\langle F \rangle$ its integrated value over
momentum space. For example, 
\be 
\langle G_{\rm HF}\rangle \equiv
\int
\frac{d^3k_1}{(2\pi)^3} G_{\rm HF}({\bf k}_1)~, 
\ee 
so that Eq.~(\ref{chi}) is simply written as 
\be 
\chi_{\rm RPA} = g
\langle G_{\rm RPA} \rangle~. 
\ee

The Bethe-Salpeter equation is an integral equation which can, in
principle, be solved iteratively 
\be 
G_{\rm RPA} = G_{\rm HF} +
G_{\rm HF} V_{\rm ph} G_{\rm HF} + G_{\rm HF} V_{\rm ph} G_{\rm
HF} V_{\rm ph} G_{\rm HF} + \dots \label{eqBSiter} 
\ee
Correspondingly, the RPA response function is written as 
\begin{widetext}
\be
\chi_{\rm RPA} = \chi_{\rm HF} + g \langle G_{\rm HF}(1) V_{\rm
ph}(1,2) G_{\rm HF}(2)\rangle + g \langle G_{\rm HF}(1) V_{\rm
ph}(1,2) G_{\rm HF}(2) V_{\rm ph}(2,3) G_{\rm HF}(3)\rangle+\cdots
\label{respexp} 
\ee 
\end{widetext}
The brackets imply integrations over chains of variables as shown
here.

In Ref.~\cite{sch89} an approximation was suggested by defining an
effective interaction $V_{\rm eff}(\romq,\omega,T)$ such that the
RPA response function is written as 
\be 
\chi_{\rm RPA}=\frac{\chi_{\rm HF}}{1-V_{\rm eff}\chi_{\rm HF}}~.
\label{chieff} 
\ee 
In the RPA neglecting exchange (the ring
approximation) the effective interaction does not depend on the
hole momenta ${\bf k}_1$ and ${\bf k}_2$ so that
Eq.~(\ref{chieff}) is exact if one replaces $V_{\rm eff}$ by
$V_{\rm ph}$. However, it is important to treat direct and
exchange terms on equal footing, since they are in general of the
same order of magnitude. Here, our point of view differs from
other works where the direct and exchange interactions are treated
on different approximation levels~\cite{Pace98}. We express the
effective interaction as a continued fraction 
\be 
V_{\rm eff} =
\frac{V_1}{1-\frac{\ds V_2 \chi_{\rm HF}}{\ds 1- \frac{\ds V_3
\chi_{\rm HF}}{\ds 1-\dots}}}~. \label{veff} 
\ee 
Each term $V_i$ entering this definition is deduced by expanding formally
Eqs.~(\ref{veff}) and (\ref{chieff}) in powers of 
$V_i \chi_{\rm HF}$ and identifying with Eq.~(\ref{respexp}). The explicit
expression for the first two terms are: 
\be
V_1 &=& \frac{g \langle G_{\rm HF} V_{\rm ph} 
G_{\rm HF} \rangle}{(\chi_{\rm HF})^2}\;,\nonumber \\
V_2 &=&\frac{g \langle G_{\rm HF} V_{\rm ph} G_{\rm HF} V_{\rm ph}
G_{\rm HF} \rangle} {V_1 \,\,\, (\chi_{\rm HF})^3}-V_1\;.
\label{V1V2} 
\ee

First, one can notice that the quantities $G_{\rm HF}$ and
$\chi_{\rm HF}$ are complex functions of $\romq$, $\omega$ and
$T$, and so are the $V_i$ and the effective interaction $V_{\rm
eff}$. Second, the calculations of the $V_i$ in the infinite
medium involve only products of functions, which is somewhat
easier numerically than the full calculations of response
functions where one needs to perform matrix
inversions~\cite{mar05}. 
Third, one see that $V_1$ is just the average of the full p-h
interaction over the squared free p-h Green's function.
Therefore, the continued fraction
approximation could be quite useful for calculating RPA response
functions if one checks how accurate it can be for a general
interaction like the Gogny force. This is what we shall examine in
Sec.~III.

\subsection{An analytical case: the Landau-Migdal interaction}

The convergence of the approximation can be explicitly seen in the
schematic case of a p-h interaction of the Landau-Migdal form
containing $\ell=0$ and $\ell=1$ terms: 
\be 
V_{\rm ph} = g \left\{
f_0 + f_1 \cos \theta_{12} \right\} \label{landau} 
\ee 
where for
brevity the same notation $f_i$ is used for the Landau parameters
in the four spin-isospin channels. For such an interaction the RPA
response function can be analytically calculated (see e.g.
Ref.~\onlinecite{Gar92}): 
\be 
\chi_{\rm RPA}=\frac{\chi_{\rm
HF}}{1- \bigg( f_0 + \frac{\ds f_1 \nu^2}{\ds 1 + F_1/3} \bigg)
\chi_{\rm HF}}~, \label{chiLan} 
\ee 
where $ \nu = \omega \mathrm{m}^*/(q k_\mathrm{F})$, $F_1= f_1 N_0$ 
is the dimensionless Landau parameter and
$N_0=g \mathrm{m}^* k_\mathrm{F} /(2 \pi^2)$ is the level density at the Fermi
surface, with $\mathrm{m}^*$ being the effective mass.

To compare with the continued fraction approximation, we have to
evaluate $V_{\rm eff}$ using the interaction (\ref{landau}). It is
sufficient to write explicitly the first 3 terms of the expansion
of $V_{\rm eff}$ and to obtain the complete series by recursion.
The integrations involving $G_{\rm HF}$ have to be carried out in
the Landau limit, i.e. $q=0$, but finite $\nu$. We get: 
\be
V_1 &=& f_0 + f_1 \nu^2 \\
V_2 &=& -\frac{1}{3}\frac{ f_1 F_1 \nu^2}{V_1 \chi_{\rm HF}}
\\
V_3 &=&\frac{1}{9} \frac{ f_1 F_1^2 \nu^2}{V_1 V_2 \chi^2_{\rm
HF}} - V_2 
\ee 
It is worth noticing that direct and exchange terms
have been treated on the same footing in calculating the $V_i$'s.
Of course for $f_1=0$ only $V_1$ is needed and one gets the exact result.
The effective interaction is 
\be 
V_{\rm eff} &=& V_1 + V_1 V_2
\chi_{\rm HF} +
\left( V_1 V_2 V_3 + V_1 V_2^2 \right) \chi^2_{\rm HF} + \dots \nonumber \\
&=& f_0 + f_1 \nu^2 \left\{
1 +  \left( - \frac{1}{3} F_1 \right) +
\left( - \frac{1}{3} F_1 \right)^2 + \dots \right\}
\label{series1}\nonumber \\
&=& f_0 + \frac{f_1 \nu^2}{1 +  \frac{1}{3}F_1}~. 
\label{series2}
\ee 
One can see that this $V_{\rm eff}$ leads to the exact result
(\ref{chiLan}) for the RPA response function.

\section{Results for a Gogny interaction}

In this section we apply the 
continuous fraction method to calculate response functions in
infinite symmetric matter for a realistic case, using the Gogny
effective interaction D1~\cite{Gog75}. 
We choose this parametrization because at the mean field level there
is a compensation between the direct and the density-dependent
contributions. Thus, it may be expected that the relative contribution
of the exchange term will be somehow enhanced.
The purpose is to
demonstrate the feasibility and rapid convergence of the method.
We only present results at $T=0$, for which the 
effects of the residual interaction are stronger.

The task of calculating the $V_i$'s involves carrying out
integrals over an increasing number of variables. We find
convenient to use a multipole expansion of both the HF propagator
$G_{\rm HF}$ and the p-h interaction $V_{\rm ph}$, as we did in
the numerically exact calculation of Ref.~\cite{mar05}: 
\be 
G_{\rm
HF}(q,\omega,{\bf k}_1)&=&\sum_\ell G_\ell(q,\omega,k_1)Y_{\ell
0}(1)~,
\nonumber \\
V_{\rm ph}(q,{\bf k}_1,{\bf k}_2)&=&\sum_{\ell,m} v_\ell(q,
k_1,k_2)Y^*_{\ell m}(1)Y_{\ell m}(2)~. 
\ee 
This allows to get rid
of all integrations over angles and we are left with only
integrals over the absolute values of momenta. For instance, we have 
\be
V_1=\frac{g}{(\chi_{\rm HF})^2}\sum_\ell \langle G_{\ell} v_{\ell}
G_{\ell} \rangle \,~, 
\ee 
where the integrals implicit in the
brackets refer now to the moduli $k_i$. Similar expressions can be
obtained for other $V_i$'s.

\begin{figure}[h]
\includegraphics[scale=0.3,clip]{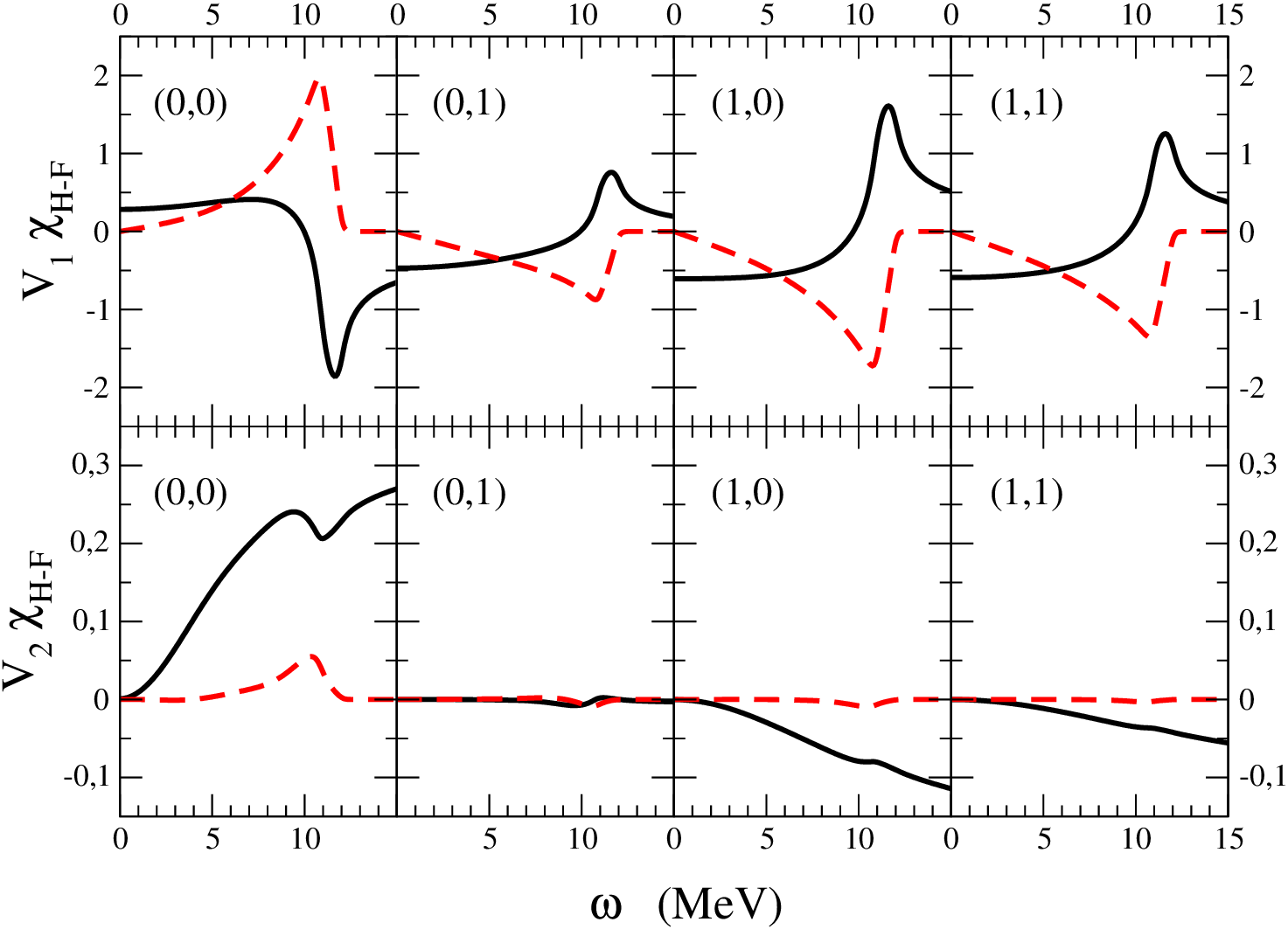}
\caption{(Color online) Real (solid line) and imaginary (dashed
line) part of $V_{1}\chi_{\rm HF}$ (top row) and $V_{2}\chi_{\rm
HF}$ (bottom row) for D1 interaction in nuclear matter at
saturation density $\rho_0$. The transferred momentum is $q$=27
MeV. The $(S,T)$ channels are shown in each panel.} \label{fig1}
\end{figure}

We can have an idea of the convergence rate by comparing the
functions $V_1 \chi_{\rm HF}$ and $V_2 \chi_{\rm HF}$. This is
shown in Fig.~\ref{fig1} for the case of a momentum transfer
$q$=27 MeV.
 Notice that the scale used
to plot $V_2 \chi_{\rm HF}$ is about a factor of ten larger than
that of $V_1 \chi_{\rm HF}$. It can be seen that the imaginary
parts of $V_2 \chi_{\rm HF}$ are close to zero for the four
spin-isospin channels. The real parts are generally small compared
to 1, but the situation seems less favorable in the channel
$(S,T)=(0,0)$. From the behavior shown in Fig.1 one can expect a
rapid convergence of the calculated
responses already at the level of $V_2$, although perhaps %for the
slower in the case of the $(0,0)$ channel.

\begin{figure}[h]
\includegraphics[scale=0.3,clip]{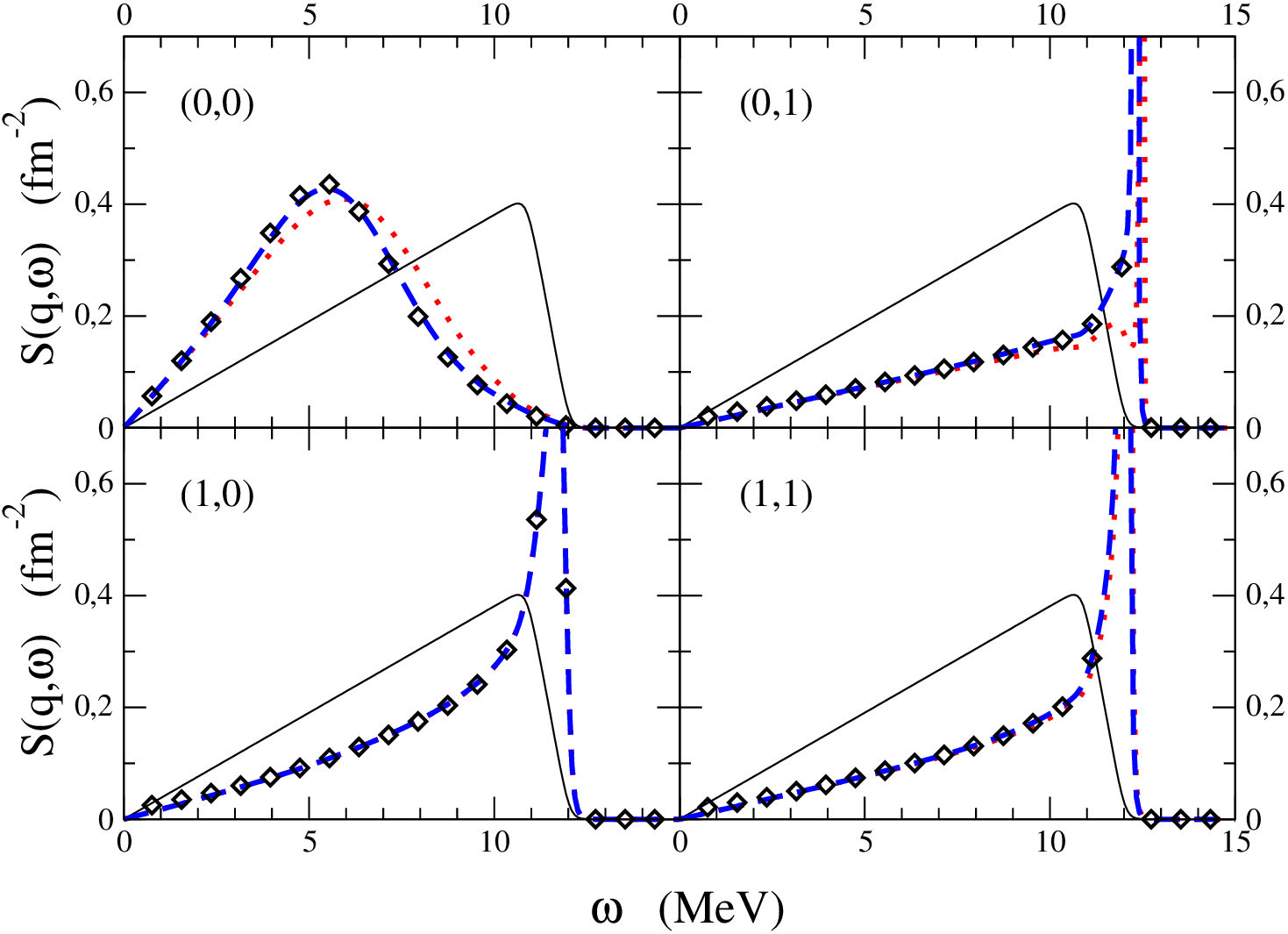}
\caption{(Color online) RPA Strength function (open diamonds)
compared with continuous fraction approximation (1st order: dotted
line, 2nd order: dashed line) calculated with Gogny D1 interaction
in
symmetric nuclear matter, at saturation density $\rho_0$ and
momentum transfer $q$=27~MeV. The thin lines represent the
uncorrelated HF strengths.} \label{fig2}
\end{figure}

\begin{figure}[h]
\includegraphics[scale=0.3,clip]{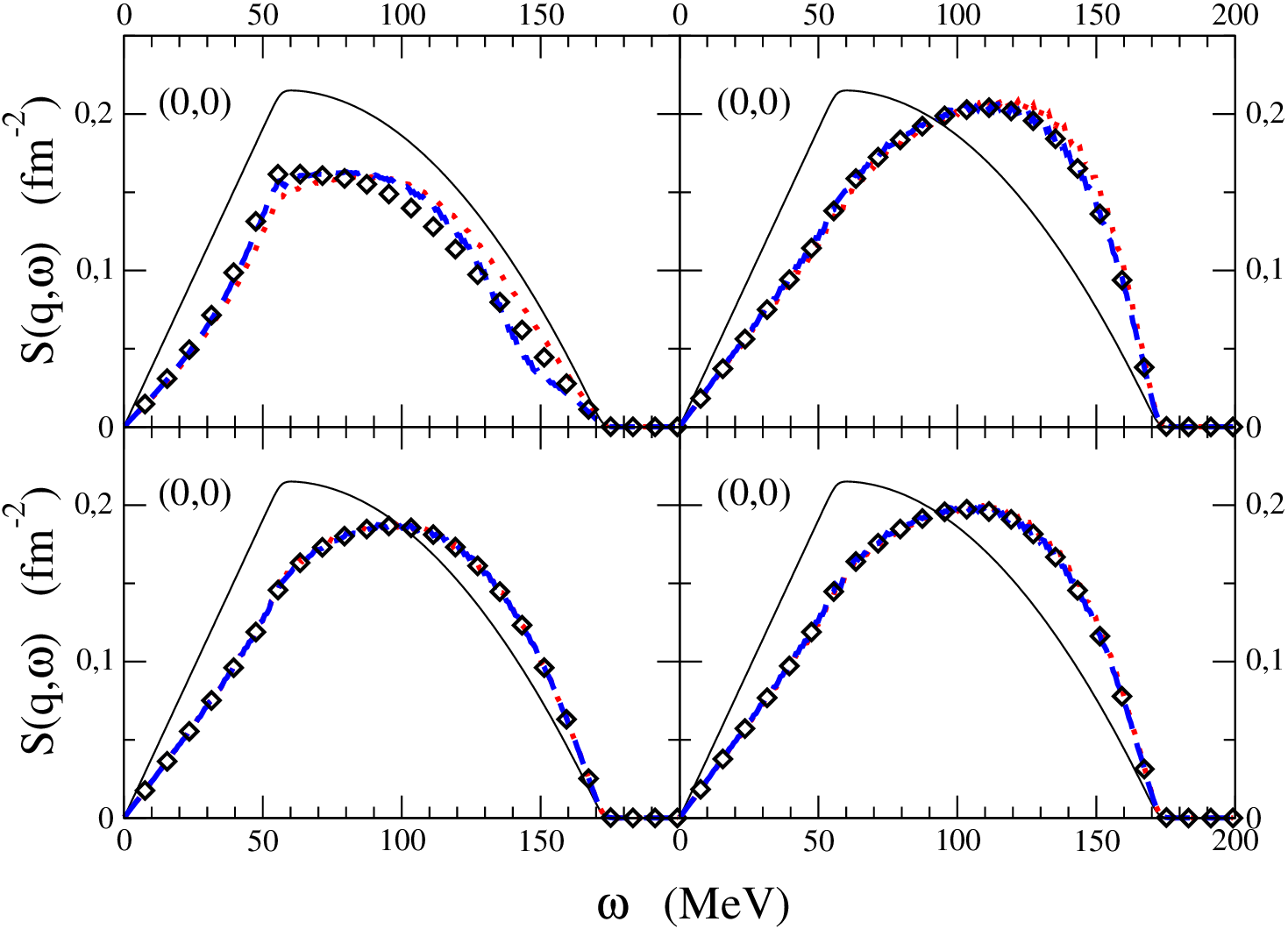}
\caption{(Color online) Same as Fig.\ref{fig2}, for $q$=270 MeV.}
\label{fig3}
\end{figure}

We now examine the strength functions 
\be
S(q,\omega)=-\frac{1}{\pi}{\rm Im}\chi(q,\omega) 
\ee 
obtained at
various levels of approximation, as compared with the direct
numerical solution of Eq.~(\ref{eqBS}) presented in
Ref.~\cite{mar05}. In Figs.~\ref{fig2}-\ref{fig3} we show the RPA
strength functions for two values of the momentum transfer, at
about $k_\mathrm{F}/10$ and $k_\mathrm{F}$. The first order gives a reliable
description of the strentgh function for all channels except
$(0,0)$ as expected from the previous analysis. For the $(0,0)$
channel it is necessary to include the second order. Notice that
the agreement is independent of the value of $q$, as no expansion
in powers of $q$ has been done. Indeed, as it can be seen in
Eq.~(\ref{veff}) the convergence of the approximation for the
effective interaction does not rely on $q$ but on the functions
$V_{i} \chi_{\rm HF}$.

\begin{figure}[h]
\includegraphics[scale=0.4,clip]{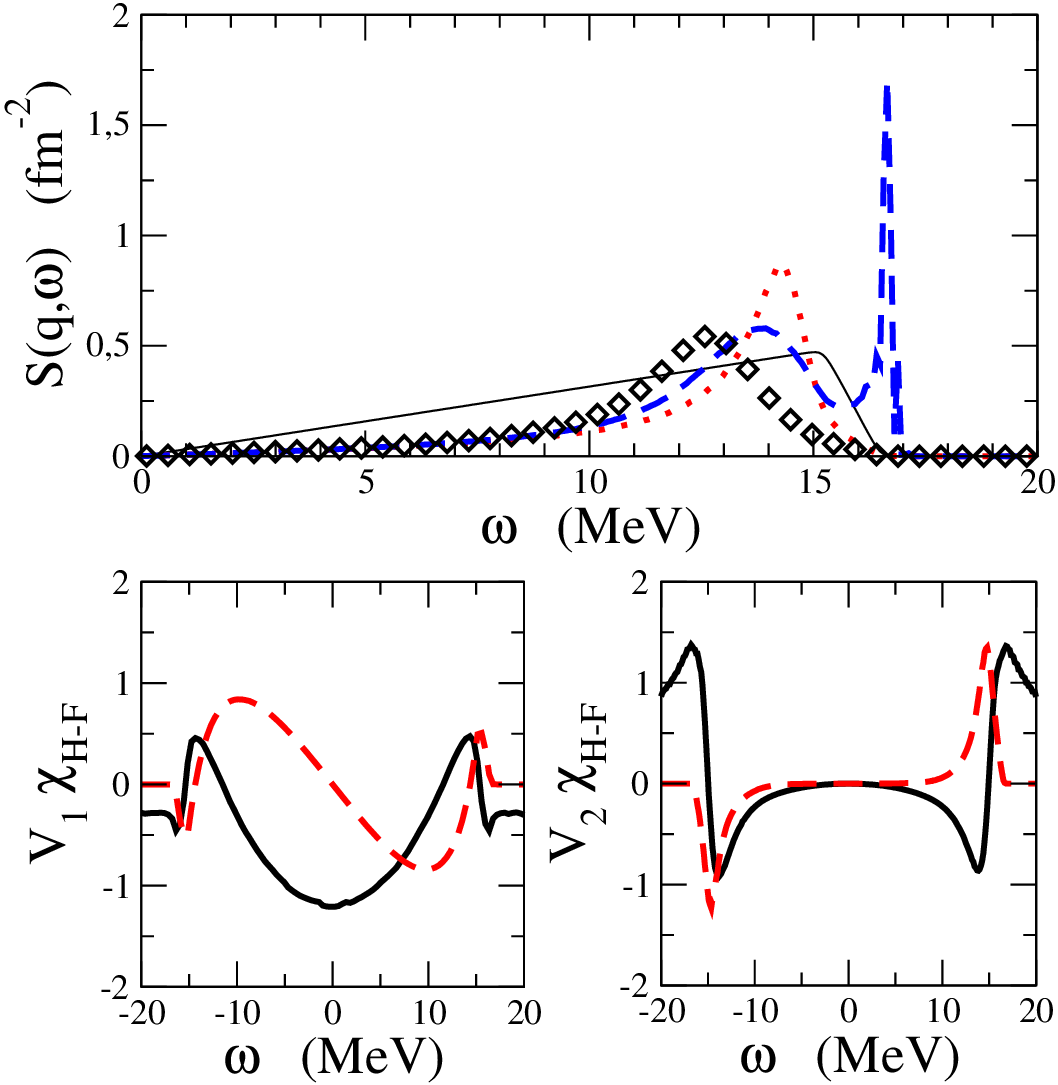}
\caption{(Color online) Same as Figs.~\ref{fig1} and~\ref{fig2}
for the channel $(0,0)$ and for $\rho= 2 \rho_0$.} \label{fig4}
\end{figure}

 However, as the density increases the convergence is
deteriorating. 
 In Fig.~\ref{fig4} are plotted the strength functions
$S^{(0,0)}$ and the functions $V_{1} \chi_{\rm HF}$, $V_{2}
\chi_{\rm HF}$ at density $\rho=2\rho_0$ in the $(0,0)$
spin-isospin channel. It can be seen that $V_2 \chi_{\rm HF}$ is
no longer small as compared to $V_{1} \chi_{\rm HF}$ and
consequently, a reliable strength function should require at least
the inclusion of third order terms in the effective interaction.
On the other hand, for densities smaller than $\rho_0$ the
approximation $V_{\rm eff} = V_1 \chi_{\rm HF}$ is sufficient to
get accurate results. Of course, the specific
convergence found in each channel $(S,T)$
depends on the specific interaction used.

Let us remind that the present approximation is not related to the
relative importance of the direct and exchange contributions to
the particle-hole interaction. Had the exchange term be small as
compared to the direct one, a good approximation for the response
function could be obtained by treating exactly the contribution of
the latter and using some approximation for the contribution of
the former term. This is not the case for the D1 interaction, as
it can be seen in Table~1. The explicit expressions of these terms
are given in Ref.~\cite{mar05}. As the exchange term depends on
momenta $k_1$, $k_2$ and their relative angle, in the Table are
plotted the monopole contributions at the Fermi surface
($k_1=k_2=k_\mathrm{F}$), for two values of $q$. 

\begin{table}[h]
\begin{tabular}{|c|c|c|c|c|c|c|c|c|}
\hline
(S,T) channel &\multicolumn{2}{c|}{(0,0)} &\multicolumn{2}{c|}{(0,1)}
&\multicolumn{2}{c|}{(1,0)} &\multicolumn{2}{c|}{(1,1)}\\[0.15cm]
\hline
$q$  (MeV) & 27 & 135 & 27 & 135 & 27 & 135 & 27 & 135 \\[0.15cm]
\hline
$v_{\ell=0}^{(D)}(q)$ & 885 & 1129 & -363 & -459 & 845 & 798 & -46 & -146\\[0.15cm]
\hline
$v_{\ell=0}^{(E)}(k_{1,2}=k_\mathrm{F})$ & -1147 & -1147 & 917 & 917 & -420 & -420 & 583 & 583 \\[0.15cm]
\hline
\end{tabular}
\caption{Direct $(D)$ and exchange terms $(E)$ in MeV.fm$^{-3}$
of the D1 p-h interaction in nuclear
matter, for $\rho=\rho_0$ and angular momentum $\ell$=0.}
\label{tab2}
\end{table}

\section{Conclusions}

We have examined the efficiency of the continuous fraction method
for calculating RPA response functions in infinite nuclear matter
using a typical finite range effective force. This issue originates from
the need of having self-consistent theoretical predictions of
nuclear responses calculated with realistic interactions.

We have found that, with the Gogny interaction D1 the continued
fraction method is very efficient and the exact RPA response
functions in the 4 spin-isospin channels are well reproduced
already at first order. This is true when the nuclear density is
of the order of, or less than the saturation density value. At
higher densities it becomes necessary to include second and higher
order terms. The rate of convergence is controlled by the decrease
of the terms of successive orders $V_i\chi_{\rm HF}$. In our
expansion the direct and exchange interactions are always treated
on equal footing.
This is important since in the nuclear case usually there occurs a
strong cancellation of two large numbers, see Table~I.

The encouraging results obtained in infinite nuclear matter open
the way to important developments. For example, the continuous
fraction method for response functions provides a simpler way to
evaluate the propagation of neutrinos in dense matter such as
inside neutron stars. The accuracy of results is under control by
the rate of decrease of the successive terms $V_i\chi_{\rm HF}$.
In finite nuclei, response functions can be calculated
consistently with realistic effective interactions without
diagonalizing RPA matrices of extremely large dimensions. This can
be of some advantage for studying heavy and/or deformed nuclei.

\subsection*{Acknowledgments}

This work is supported by the grant FIS2007-60133 (MEC, Spain) and
by the IN2P3(France)-CICYT(Spain) exchange program.

\end{document}